\def\be{\begin{equation}}
\def\ee{\end{equation}}
\def\ba{\begin{array}}
\def\ea{\end{array}}

\makeatletter

\newcommand{\Rmnum}[1]{\expandafter\@slowromancap\romannumeral #1@}
\makeatother
\documentclass[aps,amsmath,amssymb,amsfonts]{revtex4}
\usepackage{txfonts}
\usepackage{stmaryrd}

\usepackage{graphicx}
\usepackage{epstopdf}
\def\qed{\leavevmode\unskip\penalty9999 \hbox{}\nobreak\hfill
     \quad\hbox{\leavevmode  \hbox to.77778em{%
               \hfil\vrule   \vbox to.675em%
               {\hrule width.6em\vfil\hrule}\vrule\hfil}}
     \par\vskip3pt}

\begin{document}

\title{Multipartite entanglement of fermionic system in accelerated frames}
\author{Meng Shi, Hai-Mei Zhong and Zhu-Jun Zheng\footnote {zhengzj@scut.edu.cn}}
\affiliation{School of Mathematics, South China University of Technology, Guangzhou 510641, China}

\begin{abstract}

{\bf Abstract} We investigate the entanglement measures of tripartite W-State and GHZ-state in noninertial frame through the coordinate transformation between Minkowski and Rindler. First it is shown that all three qubits undergo in a uniform acceleration $a$ of W-State, we find that the one-tangle, two-tangle, and $\pi$-tangle decrease when the acceleration parameter $r$ increases, and the two-tangle cannot arrive to infinity of the acceleration. Next we show that the one qubit goes in a uniform acceleration $a_{1}$ and the other two undergo in a uniform acceleration $a$ of GHZ-state, we find that the two-tangle is equal to zero and $N_{B_{\Rmnum{1}}(A_{\Rmnum{1}}C_{\Rmnum{1}})}=N_{C_{\Rmnum{1}}(A_{\Rmnum{1}}B_{\Rmnum{1}})}\neq N_{A_{\Rmnum{1}}(B_{\Rmnum{1}}C_{\Rmnum{1}})}$, but one-tangle  and $\pi$-tangle never reduce to zero for any acceleration.

\end{abstract}

\maketitle

\section{Introduction}
Entanglement is one of the most important concepts in the quantum information and quantum computation\cite{1}, and the most desirable physical resources for a variety of quantum information processing tasks, for example quantum simulation\cite{M}, quantum teleportation\cite{C}. Since quantum information science and relativity theory permit us to have a profounder understanding of physical property of multipartite states in a noninertial frame\cite{2,E}. In the circumstances, entanglement in a noninertial frame gets an important theme in quantum information theory. So it is essential to show how the involved partites in an entangled state behave from the point of view of the accelerating observers.\\
\indent In a noninertial frame, Rindler coordinates are appropriate for discussing from the viewpoint of an observer moving with uniform acceleration. The Minkowski coordinates are the most suitable to describe the field form in an inertial frame. So it is necessary to have two different sets of coordinates to transfer field states in Minkowski space time to Rindler coordinates, which define two disconnected regions in Rindler space-time\cite{8,9}. We need to make the transformation between both coordinated systems. For the fermion field, we use the following transformation\cite{jj,21,23}:
$$|0_{\omega_{i}}\rangle_{M}=cosr_{i}|0_{\omega_{i}}\rangle_{\Rmnum{1}}|0_{w_{i}}\rangle_{\Rmnum{2}}+sinr_{i}|1_{w_{i}}\rangle_{\Rmnum{1}}|1_{w_{i}}\rangle_{\Rmnum{2}}\eqno{(1)}$$
and$$|1_{w_{i}}\rangle_{M}=|1_{w_{i}}\rangle_{\Rmnum{1}}|0_{w_{i}}\rangle_{\Rmnum{2}}\eqno{(2)}$$with $cosr_{i}=(e^{-2\pi\omega_{i}c/a_{i}}+1)^{-1/2}$, where the constants $\omega$, $c$ and $a$ in the exponential item stand for Fermi particle¡¯s frequency, the speed of light in vacuum and observer¡¯s acceleration, and the acceleration parameter $r$ is in the range $0\leq r\leq \frac{\pi}{4}$ for $0\leq a<\infty$.\\
\indent In the last few years, there are a lot of number of papers showing the entanglement in the accelerated model. At first, Fuentes-Schuller and Mann studied that the maximal bipartite entanglement decreased when the observers were relatively accelerated, at last the bipartite state  went to separable state when the acceleration approached infinity\cite{3}. Afterwards, Hwang et al. showed the tripartite entanglement of scalar field in noninertial frames, in this case the entanglement reduced as one of the observers was accelerated, but the entanglement never absolutely vanished when the acceleration approached infinity\cite{4}. Moradi and Amiri discussed the nonlocality and multipartite entanglement in asymptotically flat space times\cite{5}. The majority of the papers focus their investigation in two main states: GHZ-state, W-state. Jieci Wang and Jiliang Jing investigated that entanglement of a tripartite GHZ-state when one or two observer accelerated, they showed that all the one-tangles decreased when the acceleration increased and found that tripartite entanglement never vanished Whatever the acceleration was\cite{jj}. Ariadna J.Torres-Arenas, QianDong et al. showed entanglement measures of a tripartite W-State entangled system in noninertial frame when one qubit went in a uniform acceleration or the other two qubits underwent in another uniform acceleration, they illustrated the negativities in graphics and studied their dependencies on the acceleration parameters. And they also analyzed one-tangle, two-tangle and $\pi$-tangle decreased with the acceleration parameters increasing\cite{2019PLB}.\\
\indent The paper is arranged as follows. In Section $\Rmnum{2}$ we will mainly introduce entanglement measurement of tripartite states: $\pi$-tangle. In Section $\Rmnum{3}$ we will discuss W-state entanglement when three observers are at the same acceleration. And we will calculate the one-tangle, two-tangle, $\pi$-tangle. In Section $\Rmnum{4}$ we will introduce tripartite entanglement of GHZ-state when Alice will be moving in a uniform acceleration, meanwhile Bob and Charlie will be moving in other uniform acceleration. In Section $\Rmnum{5}$ we will summarize our results.
\section{Entanglement measurement of tripartite states}
For a bipartite system $\rho_{\alpha\beta}$, the concurrence\cite{V} and
the negativity\cite{G} are two entanglement measures. We merely introduce about negativity, it is defined as
$$\mathcal {N}_{\alpha\beta}=\|\rho^{T_{\alpha}}_{\alpha\beta}\| -1\eqno{(3)}$$
 and $\|\cdot\|$ represents the trace norm of a matrix, defined as: $\|\rho\|=tr\left(\sqrt{\rho\rho^{\dag}}\right)$\cite{33}. There are two entanglement measures that quantify the genuine tripartite entanglement: three-tangle\cite{V} and $\pi$-tangle\cite{26}. To simplify the calculation we merely adopt the $\pi$-tangle as the quantification of the tripartite entanglement. For any three-qubit state $|\phi\rangle_{\alpha\beta\gamma}$¡°,¡± $N_{\alpha\beta}$ is a ¡°two-tangle¡± which is the negativity of the mixed state $\rho_{\alpha\beta}=Tr_{\gamma}(|\phi\rangle_{\alpha\beta\gamma}\langle\phi|)$ and $N_{\alpha(\beta\gamma)}$ is a ¡°one-tangle¡± defined as $N_{\alpha(\beta\gamma)}=\|\rho^{T_{\alpha}}_{\alpha\beta\gamma}\|-1$¡°.¡± The partial transpose criterion provides an enough criterion for entanglement, only if at least one eigenvalue is negative then the density matrix is entangled. This is called distillable entanglement. On the other hand, $\|A\|-1$ is equal to two times of the sum of absolute values of negative eigenvalues of the matrix $A$. That is to say, for one-tangle\cite{2019PLB}:$$N_{\alpha(\beta\gamma)}=2\sum_{i=1}^N|\lambda^{(-)}_{\alpha(\beta\gamma)}|^{i}\eqno{(4)}$$ and for two-tangle $$N_{\alpha\beta}=2\sum_{i=1}^N|\lambda^{(-)}_{\alpha\beta}|^{i}\eqno{(5)}$$
where$[\lambda^{(-)}_{\alpha(\beta\gamma)}]^{i}$ and $[\lambda^{(-)}_{\alpha\beta}]^{i}$ represent all negative eigenvalues of the partial transpose matrix. Residual entanglement is defined as \cite{jj}:
$$\pi_{\alpha}=N^{2}_{\alpha(\beta\gamma)}-N^{2}_{\alpha\beta}-N^{2}_{\alpha\gamma}$$
$$\pi_{\beta}=N^{2}_{\beta(\alpha\gamma)}-N^{2}_{\beta\alpha}-N^{2}_{\beta\gamma}$$
$$\pi_{\gamma}=N^{2}_{\gamma(\alpha\beta)}-N^{2}_{\gamma\alpha}-N^{2}_{\gamma\beta}$$
In general, $\pi_{\alpha}\neq\pi_{\beta}\neq\pi_{\gamma}$, it indicates that the residual entanglement corresponding to the different focus is variant under permutations of the qubits. The $\pi$-tangle $\pi_{\alpha\beta\gamma}$ is defined as:
$$\pi_{\alpha\beta\gamma}=\frac{1}{3}(\pi_{\alpha}+\pi_{\beta}+\pi_{\gamma}).\eqno{(6)}$$
It is satisfied that $\pi_{\alpha\beta\gamma}\geq0$, and $\pi_{\alpha\beta\gamma}=0$ for product pure states. The $\pi$-tangle is shown to be a natural entanglement measure and can be extended to mixed states and general pure n-qubit states\cite{26}.
\section{Tripartite entanglement of W-state when three observers are accelerated}
 We will be considering the W-class entangled state which is composed by fermions, which in this case are three qubits with the name of Alice, Bob, Charlie. The paper\cite{2019PLB} has introduced  the change of the entanglement of the W-state with acceleration and has considered the two cases: when one qubit goes in a uniform acceleration $a$ and the others remain stationary and when two qubits undergo in a uniform acceleration and while the other is stationary. And they illustrated one-tangle, two-tangle and $\pi$-tangle in graphics. let us consider that Alice, Bob and Charlie meanwhile will be moving in a uniform acceleration $a$. The tripartite W-state is as follows and subscripts ABC are used to denote the qubits Alice, Bob and Charlie:
$$|W\rangle=\frac{1}{\sqrt{3}}(|001\rangle_{A,B,C}+|010\rangle_{A,B,C}+|100\rangle_{A,B,C})$$
We apply the Eqs $(1)$ and $(2)$ to the $|W\rangle$ state and get the following state:
\begin{eqnarray*}
|\psi\rangle_{A_{(\Rmnum{1},\Rmnum{2})}B_{({\Rmnum{1},\Rmnum{2}})}C_{(\Rmnum{1},\Rmnum{2})}}=\frac{1}{\sqrt{3}}[
cos^{2}r |000010\rangle_{A_{(\Rmnum{1},\Rmnum{2})}B_{(\Rmnum{1},\Rmnum{2})}C_{(\Rmnum{1},\Rmnum{2})}}
+sinrcosr |110010\rangle_{A_{(\Rmnum{1},\Rmnum{2})}B_{(\Rmnum{1},\Rmnum{2})}C_{(\Rmnum{1},\Rmnum{2})}}\\
+cos^{2}r |001000\rangle_{A_{(\Rmnum{1},\Rmnum{2})}B_{(\Rmnum{1},\Rmnum{2})}C_{(\Rmnum{1},\Rmnum{2})}}
+cosrsinr |111000\rangle_{A_{(\Rmnum{1},\Rmnum{2})}B_{(\Rmnum{1},\Rmnum{2})}C_{(\Rmnum{1},\Rmnum{2})}}\\
+sinrcosr |001011\rangle_{A_{(\Rmnum{1},\Rmnum{2})}B_{(\Rmnum{1},\Rmnum{2})}C_{(\Rmnum{1},\Rmnum{2})}}
+sin^{2}r |111011\rangle_{A_{(\Rmnum{1},\Rmnum{2})}B_{(\Rmnum{1},\Rmnum{2})}C_{(\Rmnum{1},\Rmnum{2})}}\\
+sinrcosr |001110\rangle_{A_{(\Rmnum{1},\Rmnum{2})}B_{(\Rmnum{1},\Rmnum{2})}C_{(\Rmnum{1},\Rmnum{2})}}
+sin^{2}r |111110\rangle_{A_{(\Rmnum{1},\Rmnum{2})}B_{(\Rmnum{1},\Rmnum{2})}C_{(\Rmnum{1},\Rmnum{2})}}\\
+cos^{2}r |100000\rangle_{A_{(\Rmnum{1},\Rmnum{2})}B_{(\Rmnum{1},\Rmnum{2})}C_{(\Rmnum{1},\Rmnum{2})}}
+cosrsinr |100011\rangle_{A_{(\Rmnum{1},\Rmnum{2})}B_{(\Rmnum{1},\Rmnum{2})}C_{(\Rmnum{1},\Rmnum{2})}}\\
+cosrsinr |101100\rangle_{A_{(\Rmnum{1},\Rmnum{2})}B_{(\Rmnum{1},\Rmnum{2})}C_{(\Rmnum{1},\Rmnum{2})}}
+sin^{2}r |101111\rangle_{A_{(\Rmnum{1},\Rmnum{2})}B_{(\Rmnum{1},\Rmnum{2})}C_{(\Rmnum{1},\Rmnum{2})}} ]
\end{eqnarray*}
where $r$ is the acceleration parameter of Alice, Bob and Charlie. Because the Rindler modes in region $\Rmnum{2}$ are inaccessible, tracing out over the second, fourth, sixth qubits, we get the following state:
$$\rho_{A_{\Rmnum{1}}B_{\Rmnum{1}}C_{\Rmnum{1}}}=\frac{1}{3}
 \left[
 \begin{matrix}
   0 &  0 & 0 & 0 &0 & 0 & 0 & 0\\
   0 &  cos^{4}r & cos^{4}r & 0 & cos^{4}r & 0 & 0 & 0\\
   0 &  cos^{4}r & cos^{4}r & 0 & cos^{4}r & 0 & 0 & 0\\
   0 &  0 & 0 & 2sin^{2}rcos^{2}r & 0 & sin^{2}rcos^{2}r & sin^{2}rcos^{2}r & 0\\
   0 &  cos^{4}r & cos^{4}r & 0 & cos^{4}r & 0 & 0 & 0\\
   0 &  0 & 0 & sin^{2}rcos^{2}r   & 0 & 2sin^{2}rcos^{2}r & sin^{2}rcos^{2}r & 0 \\
   0 &  0 & 0 & sin^{2}rcos^{2}r  & 0 & sin^{2}rcos^{2}r & 2sin^{2}rcos^{2}r & 0 \\
   0 &  0 & 0 & 0 &0 & 0 & 0 & 3sin^{4}r\\
  \end{matrix}
   \right]
$$
Then we get the partial transpose subsystem A as following:
$$\rho^{T_{A_{\Rmnum{1}}}}_{A_{\Rmnum{1}}B_{\Rmnum{1}}C_{\Rmnum{1}}}=\frac{1}{3}
 \left[
 \begin{matrix}
   0 &  0 & 0 & 0 &0 &  cos^{4}r &  cos^{4}r & 0\\
   0 &  cos^{4}r & cos^{4}r & 0 & 0 & 0 & 0 &sin^{2}rcos^{2}r \\
   0 &  cos^{4}r & cos^{4}r & 0 & 0 & 0 & 0 & sin^{2}rcos^{2}r\\
   0 &  0 & 0 & 2sin^{2}rcos^{2}r & 0 & 0 & 0 & 0\\
   0 &  0 & 0 & 0 & cos^{4}r & 0 & 0 & 0\\
   cos^{4}r &  0 & 0 & 0   & 0 & 2sin^{2}rcos^{2}r & sin^{2}rcos^{2}r & 0 \\
   cos^{4}r &  0 & 0 & 0  & 0 & sin^{2}rcos^{2}r & 2sin^{2}rcos^{2}r & 0 \\
   0 & sin^{2}rcos^{2}r  & sin^{2}rcos^{2}r & 0 &0 & 0 & 0 & 3sin^{4}r\\
  \end{matrix}
   \right]
$$
Similarly,
$$\rho^{T_{B_{\Rmnum{1}}}}_{A_{\Rmnum{1}}B_{\Rmnum{1}}C_{\Rmnum{1}}}=\frac{1}{3}
 \left[
 \begin{matrix}
   0 &  0 & 0 &  cos^{4}r &0 & 0 &  cos^{4}r & 0\\
   0 &  cos^{4}r & 0 & 0 & cos^{4}r & 0 & 0 &sin^{2}rcos^{2}r \\
   0 &  0 & cos^{4}r & 0 & 0 & 0 & 0 & 0\\
   cos^{4}r  &  0 & 0 & 2sin^{2}rcos^{2}r & 0 & 0& sin^{2}rcos^{2}r & 0\\
   0 &  cos^{4}r & 0& 0 & cos^{4}r & 0 & 0 &  sin^{2}rcos^{2}r\\
   0 &  0 & 0 & 0  & 0 & 2sin^{2}rcos^{2}r & 0 & 0 \\
   cos^{4}r &  0 & 0 & sin^{2}rcos^{2}r  & 0 & 0 & 2sin^{2}rcos^{2}r & 0 \\
   0 & sin^{2}rcos^{2}r  & 0 & 0 &sin^{2}rcos^{2}r & 0 & 0 & 3sin^{4}r\\
  \end{matrix}
   \right]
$$
\begin{figure}
\renewcommand{\figurename}{Fig.}
  \centering
  \includegraphics[width=7cm]{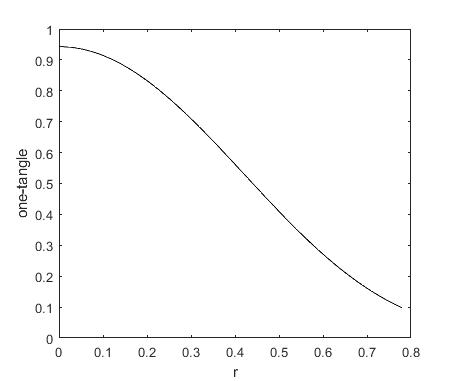}\\
  \caption{one-tangle of W-state is a function of the acceleration parameter $r$.}
\label{Fig.1}
\end{figure}
\begin{figure}
\renewcommand{\figurename}{Fig.}
  \centering
  \includegraphics[width=7cm]{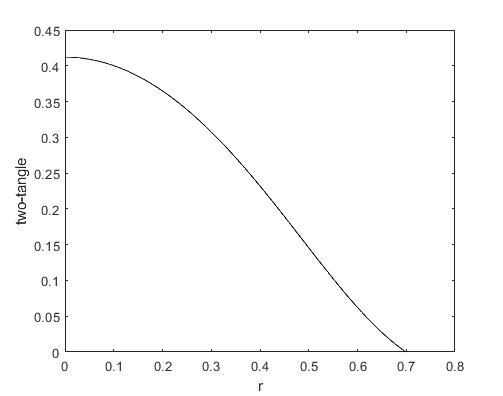}\\
  \caption{two-tangle is a function of the acceleration parameter $r$}
\label{Fig.2}
\end{figure}
 \begin{figure}
\renewcommand{\figurename}{Fig.}
  \centering
  \includegraphics[width=7cm]{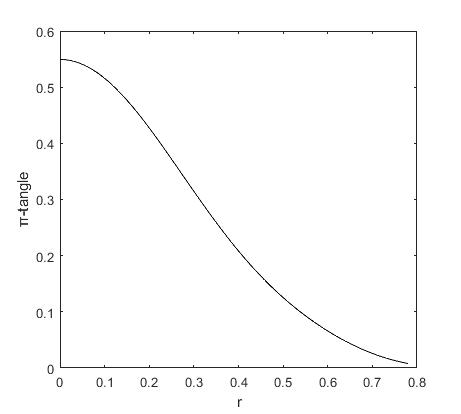}\\
  \caption{$\pi$-tangle as a function of the acceleration parameter $r$}
\label{Fig.3}
\end{figure}
$$\rho^{T_{C_{\Rmnum{1}}}}_{A_{\Rmnum{1}}B_{\Rmnum{1}}C_{\Rmnum{1}}}=\frac{1}{3}
 \left[
 \begin{matrix}
   0 &  0 & 0 &  cos^{4}r &0 & cos^{4}r & 0  & 0\\
   0 &  cos^{4}r & 0 & 0 & 0& 0 & 0 &0 \\
   0 &  0 & cos^{4}r & 0 & cos^{4}r & 0 & 0 & sin^{2}rcos^{2}r\\
   cos^{4}r  &  0 & 0 & 2sin^{2}rcos^{2}r & 0 & sin^{2}rcos^{2}r& 0 & 0\\
   0 &  0&cos^{4}r& 0 & cos^{4}r & 0 & 0 &  sin^{2}rcos^{2}r\\
   cos^{4}r &  0 & 0 &  sin^{2}rcos^{2}r & 0 & 2sin^{2}rcos^{2}r & 0 & 0 \\
   0&  0 & 0 & 0 & 0 & 0 & 2sin^{2}rcos^{2}r & 0 \\
   0 & 0  &sin^{2}rcos^{2}r & 0 &sin^{2}rcos^{2}r & 0 & 0 & 3sin^{4}r\\
  \end{matrix}
   \right]$$
We employ use Eqs$(4)$, and obtain the required one-tangle. And we find that: $$N_{A_{\Rmnum{1}}(B_{\Rmnum{1}}C_{\Rmnum{1}})}(r)= N_{B_{\Rmnum{1}}(A_{\Rmnum{1}}C_{\Rmnum{1}})}(r)=N_{C_{\Rmnum{1}}(A_{\Rmnum{1}}B_{\Rmnum{1}})}(r)=-2 \left(\frac{1}{2} cos^{2}(r)-\frac{1}{2} cos^{4}(r)-\frac{1}{6}cos^{2}(r)\sqrt{17cos^{4}(r)-18 cos^{2}(r)+9}\right)$$ In the special cases, we have $N_{A_{\Rmnum{1}}(B_{\Rmnum{1}}C_{\Rmnum{1}})}(0)= 0.9428$ , $N_{A_{\Rmnum{1}}(B_{\Rmnum{1}}C_{\Rmnum{1}})}(\frac{\pi}{4})= 0.0971$. \\
\indent Then we will get two-tangle, corresponding density matrices are given by:
$$\rho^{T_{A_{\Rmnum{1}}}}_{A_{\Rmnum{1}}B_{\Rmnum{1}}}=
\frac{1}{3}
 \left[
 \begin{matrix}
   cos^{4}r &  0 & 0 &  cos^{4}r+sin^{2}rcos^{2}r\\
   0 &  cos^{4}r+2sin^{2}rcos^{2}r & 0 & 0  \\
   0 &  0 & cos^{4}r+2sin^{2}rcos^{2}r & 0 \\
   cos^{4}r+sin^{2}rcos^{2}r  &  0 & 0 & 3sin^{4}r+2sin^{2}rcos^{2}r\\
 \end{matrix}
\right]$$
On the other hand, it is not difficult to find that: $$\rho^{T_{A_{\Rmnum{1}}}}_{A_{\Rmnum{1}}B_{\Rmnum{1}}}=\rho^{T_{A_{\Rmnum{1}}}}_{A_{\Rmnum{1}}C_{\Rmnum{1}}}=\rho^{T_{B_{\Rmnum{1}}}}_{B_{\Rmnum{1}}A_{\Rmnum{1}}}=\rho^{T_{B_{\Rmnum{1}}}}_{B_{\Rmnum{1}}C_{\Rmnum{1}}}=\rho^{T_{C_{\Rmnum{1}}}}_{C_{\Rmnum{1}}A_{\Rmnum{1}}}=\rho^{T_{C_{\Rmnum{1}}}}_{C_{\Rmnum{1}}B_{\Rmnum{1}}}\\$$ And we calculate the two-tangle, they are equal to
 $$-2 \left(\frac{1}{3} cos^{4}(r)- \frac{2}{3}cos^{2}(r)- \frac{1}{6} \sqrt{20cos^{4}(r)-24cos^{2}(r)+9}+ \frac{1}{2}\right)$$
In particular, we find that $N_{A_{\Rmnum{1}}B_{\Rmnum{1}}}(0)= 0.4120, N_{A_{\Rmnum{1}}B_{\Rmnum{1}}}(0.6970)=0$. And we also found that the two-tangle cannot arrive to infinity of the acceleration. Meanwhile we obtain the $\pi$-tangle, apply the Eqs $(6)$:$$\pi_{A_{\Rmnum{1}}B_{\Rmnum{1}}C_{\Rmnum{1}}}=\frac{1}{3}\left(3 N^{2}_{A_{\Rmnum{1}}(B_{\Rmnum{1}}C_{\Rmnum{1}})}-6 N^{2}_{A_{\Rmnum{1}}B_{\Rmnum{1}}}\right)$$
According to the data we can get the following graph, the properties of  the one-tangle, two-tangle and $\pi$-tangle of $\rho_{A_{\Rmnum{1}}B_{\Rmnum{1}}C_{\Rmnum{1}}}$ are shown in Fig. 1, Fig. 2, Fig. 3. As can be seen from the graph, one-tangle, two-tangle and $\pi$-tangle decrease with the acceleration parameter $r$ increasing, the one-tangles never disappear when the acceleration approaches infinity. On the contrary, the two-tangles will vanish as the acceleration is close to $0.7$. And the $\pi$-tangle can not reduce to zero with the acceleration approaching infinity.
\section{GHZ-state entanglement when three  observer are accelerated}
The paper\cite{jj} has discussed  both the bipartite and tripartite entanglement of tripartite GHZ-state in the noninertial frame when one or two observers are accelerated. It is shown that all the one-tangles decrease as the acceleration increases and the $\pi$-tangle of the two-observer-accelerated case decreases much quicker than that of the one-observer-accelerated case. Now we will consider that Alice will be moving in a uniform acceleration $a_{1}$, meanwhile Bob and Charlie will be moving in a uniform acceleration $a$, and the corresponding acceleration parameters are $r_{a}$, $r$. We will be considering the GHZ-state, which is three qubits:
$$|GHZ\rangle_{A,B,C}=\frac{1}{\sqrt{2}}(|000\rangle_{A,B,C}+|111\rangle_{A,B,C})$$
We apply the Eqs $(1)$ and $(2)$ to the GHZ-state and get the following state:
\begin{eqnarray*}
|\varphi\rangle_{A_{(\Rmnum{1},\Rmnum{2})}B_{({\Rmnum{1},\Rmnum{2}})}C_{(\Rmnum{1},\Rmnum{2})}}=\frac{1}{\sqrt{2}}[
  cos^{2}rcosr_{a} |000000\rangle_{A_{(\Rmnum{1},\Rmnum{2})}B_{(\Rmnum{1},\Rmnum{2})}C_{(\Rmnum{1},\Rmnum{2})}}
  +cos^{2}rsinr_{a} |110000\rangle_{A_{(\Rmnum{1},\Rmnum{2})}B_{(\Rmnum{1},\Rmnum{2})}C_{(\Rmnum{1},\Rmnum{2})}}\\
  +cosrsinrcosr_{a} |000011\rangle_{A_{(\Rmnum{1},\Rmnum{2})}B_{(\Rmnum{1},\Rmnum{2})}C_{(\Rmnum{1},\Rmnum{2})}}
  +cosrsinrsinr_{a} |110011\rangle_{A_{(\Rmnum{1},\Rmnum{2})}B_{(\Rmnum{1},\Rmnum{2})}C_{(\Rmnum{1},\Rmnum{2})}}\\
  +sinrcosrcosr_{a} |001100\rangle_{A_{(\Rmnum{1},\Rmnum{2})}B_{(\Rmnum{1},\Rmnum{2})}C_{(\Rmnum{1},\Rmnum{2})}}
  +cosrsinrsinr_{a}|111100\rangle_{A_{(\Rmnum{1},\Rmnum{2})}B_{(\Rmnum{1},\Rmnum{2})}C_{(\Rmnum{1},\Rmnum{2})}}\\
  +sin^{2}rcosr_{a}  |001111\rangle_{A_{(\Rmnum{1},\Rmnum{2})}B_{(\Rmnum{1},\Rmnum{2})}C_{(\Rmnum{1},\Rmnum{2})}}
  +sin^{2}rsinr_{a}|111111\rangle_{A_{(\Rmnum{1},\Rmnum{2})}B_{(\Rmnum{1},\Rmnum{2})}C_{(\Rmnum{1},\Rmnum{2})}}\\
  +|101010\rangle_{A_{(\Rmnum{1},\Rmnum{2})}B_{(\Rmnum{1},\Rmnum{2})}C_{(\Rmnum{1},\Rmnum{2})}}]
\end{eqnarray*}
where $r_{a}$ is the acceleration parameter of Alice, $r$ is the acceleration parameter of Bob and Charlie. Because the Rindler modes in region $\Rmnum{2}$ are inaccessible, we trace out over the second, fourth, sixth qubits, and then get the partial transpose subsystem A as following: $\sigma^{T_{A_{\Rmnum{1}}}}_{A_{\Rmnum{1}}B_{\Rmnum{1}}C_{\Rmnum{1}}}$=
$$\frac{1}{2}
 \left[
 \begin{matrix}
  cos^{4}rcos^{2}r_{a} & 0 & 0 & 0 &0 & 0 & 0 & 0\\
  0 & cos^{2}rsin^{2}rcos^{2}r_{a} & 0 & 0 & 0 & 0 & 0 & 0\\
  0 & 0 &  cos^{2}rsin^{2}rcos^{2}r_{a} & 0 & 0 & 0 & 0 & 0\\
  0 & 0 & 0 & sin^{4}rcos^{2}r_{a}&  cos^{2}rcosr_{a} & 0& 0 & 0\\
  0 & 0& 0 & cos^{2}rcosr_{a} & cos^{4}rsin^{2}r_{a} & 0 & 0 & 0\\
  0 & 0 & 0 & 0  & 0 & sin^{2}rcos^{2}rsin^{2}r_{a} & 0 & 0 \\
  0 & 0 & 0 & 0& 0 & 0 & sin^{2}rcos^{2}rsin^{2}r_{a} & 0 \\
  0 & 0 & 0 & 0 &0 & 0 & 0 & 1+sin^{4}rsin^{2}r_{a}\\
  \end{matrix}
   \right]
$$
Similarly,\\
$\sigma^{T_{B_{\Rmnum{1}}}}_{A_{\Rmnum{1}}B_{\Rmnum{1}}C_{\Rmnum{1}}}$=
$$\frac{1}{2}
 \left[
 \begin{matrix}
  cos^{4}rcos^{2}r_{a} & 0 & 0 & 0 &0 & 0 & 0 & 0\\
  0 & cos^{2}rsin^{2}rcos^{2}r_{a} & 0 & 0 & 0 & 0 & 0 & 0\\
  0 & 0 &  cos^{2}rsin^{2}rcos^{2}r_{a} & 0 & 0 &cos^{2}rcosr_{a} & 0 & 0\\
  0 & 0 & 0 & sin^{4}rcos^{2}r_{a}& 0& 0& 0 & 0\\
  0 & 0& 0 &0 & cos^{4}rsin^{2}r_{a} & 0 & 0 & 0\\
  0 & 0 & cos^{2}rcosr_{a} & 0  & 0 & sin^{2}rcos^{2}rsin^{2}r_{a} & 0 & 0 \\
  0 & 0 & 0 & 0& 0 & 0 & sin^{2}rcos^{2}rsin^{2}r_{a} & 0 \\
  0 & 0 & 0 & 0 &0 & 0 & 0 & 1+sin^{4}rsin^{2}r_{a}\\
  \end{matrix}
   \right]
$$

$\sigma^{T_{C_{\Rmnum{1}}}}_{A_{\Rmnum{1}}B_{\Rmnum{1}}C_{\Rmnum{1}}}$=
$$\frac{1}{2}
 \left[
 \begin{matrix}
  cos^{4}rcos^{2}r_{a} & 0 & 0 & 0 &0 & 0 & 0 & 0\\
  0 & cos^{2}rsin^{2}rcos^{2}r_{a} & 0 & 0 & 0 & 0 & cos^{2}rcosr_{a} & 0\\
  0 & 0 &  cos^{2}rsin^{2}rcos^{2}r_{a} & 0 & 0 &0 & 0 & 0\\
  0 & 0 & 0 & sin^{4}rcos^{2}r_{a}& 0& 0& 0 & 0\\
  0 & 0& 0 &0 & cos^{4}rsin^{2}r_{a} & 0 & 0 & 0\\
  0 & 0 & 0 & 0  & 0 & sin^{2}rcos^{2}rsin^{2}r_{a} & 0 & 0 \\
  0 & cos^{2}rcosr_{a}& 0 & 0& 0 & 0 & sin^{2}rcos^{2}rsin^{2}r_{a} & 0 \\
  0 & 0 & 0 & 0 &0 & 0 & 0 & 1+sin^{4}rsin^{2}r_{a}\\
  \end{matrix}
   \right]
$$
The next we make use of Eqs $(4)$ and $(5)$ to calculate all required negativities one-tangle and two-tangle, and $\pi$-tangle. Because of the obtained calculations are not easy to express in a short term, we do not try to write them out explicitly and find that $N_{B_{\Rmnum{1}}(A_{\Rmnum{1}}C_{\Rmnum{1}})}=N_{C_{\Rmnum{1}}(A_{\Rmnum{1}}B_{\Rmnum{1}})}\neq N_{A_{\Rmnum{1}}(B_{\Rmnum{1}}C_{\Rmnum{1}})}.$
\begin{figure}
\renewcommand{\figurename}{Fig.}
  \centering
  \includegraphics[width=7cm]{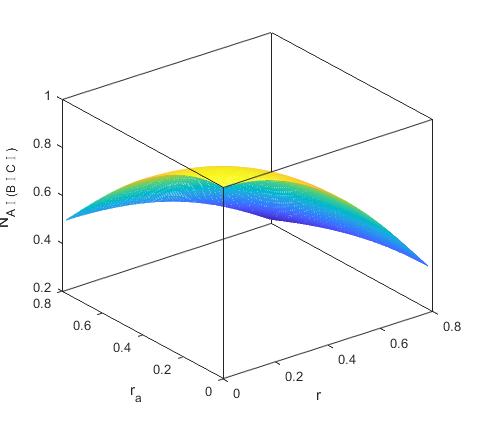}\\
  \caption{$N_{A_{\Rmnum{1}}(B_{\Rmnum{1}}C_{\Rmnum{1}})}$ as a function of the acceleration parameter $r$ and $r_a$}
\label{Fig.1}
\end{figure}

\begin{figure}
\renewcommand{\figurename}{Fig.}
  \centering
  \includegraphics[width=7cm]{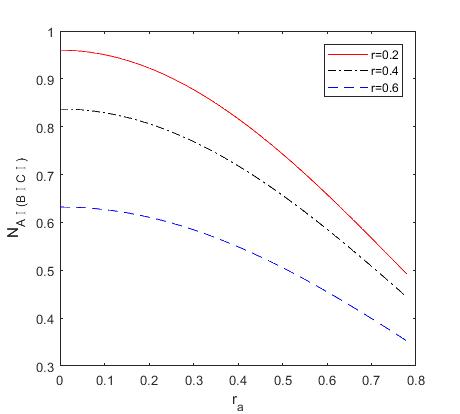}\\
  \caption{$N_{A_{\Rmnum{1}}(B_{\Rmnum{1}}C_{\Rmnum{1}})}$ as a function of the acceleration parameter $r_a$}
\label{Fig.1}
\end{figure}

\begin{figure}
\renewcommand{\figurename}{Fig.}
  \centering
  \includegraphics[width=7cm]{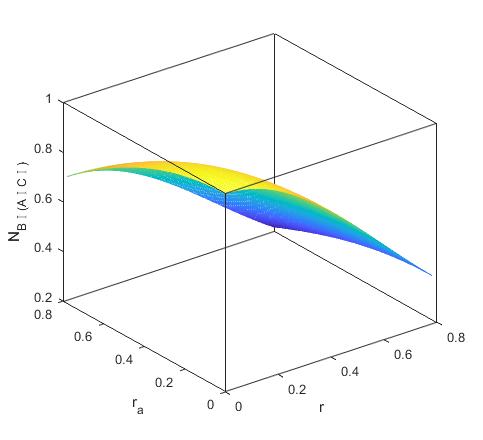}\\
  \caption{$N_{B_{\Rmnum{1}}(A_{\Rmnum{1}}C_{\Rmnum{1}})}$ as a function of the acceleration parameter $r$ and $r_a$}
\label{Fig.1}
\end{figure}

\begin{figure}
\renewcommand{\figurename}{Fig.}
  \centering
  \includegraphics[width=7cm]{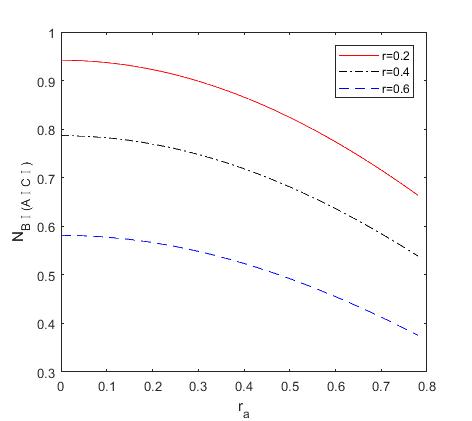}\\
  \caption{$N_{B_{\Rmnum{1}}(A_{\Rmnum{1}}C_{\Rmnum{1}})}$ as a function of the acceleration parameter $r_a$}
\label{Fig.1}
\end{figure}

\begin{figure}
\renewcommand{\figurename}{Fig.}
  \centering
  \includegraphics[width=7cm]{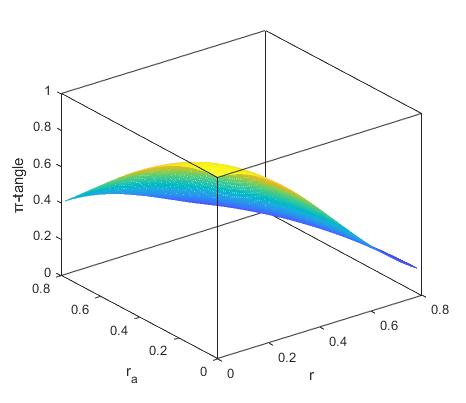}\\
  \caption{$\pi-tangle$ as a function of the acceleration parameter $r$ and $r_a$}
\label{Fig.1}
\end{figure}

\begin{figure}
\renewcommand{\figurename}{Fig.}
  \centering
  \includegraphics[width=7cm]{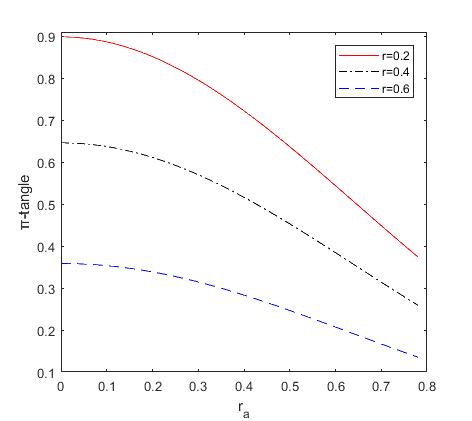}\\
  \caption{$\pi-tangle$ as a function of the acceleration parameter $r_a$}
\label{Fig.1}
\end{figure}
By computing the matrix, we find that all the two-tangle are equal to zero. Meanwhile we obtain the $\pi$-tangle, apply the Eqs $(6)$:
$$\pi_{A_{\Rmnum{1}}B_{\Rmnum{1}}C_{\Rmnum{1}}}=\frac{1}{3}(N^{2}_{B_{\Rmnum{1}}(A_{\Rmnum{1}}C_{\Rmnum{1}})}+N^{2}_{C_{\Rmnum{1}}(A_{\Rmnum{1}}B_{\Rmnum{1}})}+N^{2}_{A_{\Rmnum{1}}(B_{\Rmnum{1}}C_{\Rmnum{1}})})$$
As we can see from the graph, one-tangle and $\pi$-tangle decrease when Alice will be moving in a uniform acceleration $a$ and meanwhile Bob and Charlie will be moving in a steady acceleration, but they never reduce to zero with the acceleration approaching infinity.
\section{conclusion}
In this paper, the behavior of entanglement of fermionic GHZ-state and W-state is
investigated when three observers are all in the accelerated frames. As for the W-state, we have studied negativities for one-tangle, two-tangle and $\pi$-tangle when Alice, Bob and Charlie meanwhile will be moving in a uniform acceleration $a$, the negativities dependence on the acceleration parameter $r$ through graphics, negativities will decrease with acceleration increasing. The one-tangles never disappear when the acceleration approaches infinity. On the contrary, the two-tangles will vanish as the acceleration is close to $0.7$. We also find that $N_{A_{\Rmnum{1}}(B_{\Rmnum{1}}C_{\Rmnum{1}})}=N_{B_{\Rmnum{1}}(A_{\Rmnum{1}}C_{\Rmnum{1}})}=N_{C_{\Rmnum{1}}(A_{\Rmnum{1}}B_{\Rmnum{1}})}$.  And the $\pi$-tangle can not reduce to zero with the acceleration approaching infinity. Besides we also have showed negativities of GHZ-state. It is showed that $N_{B_{\Rmnum{1}}(A_{\Rmnum{1}}C_{\Rmnum{1}})}=N_{C_{\Rmnum{1}}(A_{\Rmnum{1}}B_{\Rmnum{1}})}$, and all two-tangle are equal to zero. we also showed that the one-tangle and $\pi$-tangle will decrease when Alice will be moving in a uniform acceleration  parameter $r_a$, meanwhile Bob and Charlie will be moving in a stationary acceleration, but they never reach zero. Therefore, further investigation by using
the results in this paper will help us better understand genuine multipartite entanglement in noninertial frame.

\section*{Acknowledgments}
 This work is supported by the NSFC 11571119.

\end{document}